\documentclass[aps,pre,twocolumn,superscriptaddress,groupedaddress,showpacs,amsmath,amssymb,]{revtex4}
\usepackage{epsfig}

\begin{document}

\title{Universal free energy distribution in the critical point of a random Ising ferromagnet}

\author{Victor Dotsenko}\email{dotsenko@lptl.jussieu.fr}
\affiliation{LPTMC, Universit\'e Paris VI,  75252 Paris, France}
\affiliation{L.D.\ Landau Institute for Theoretical Physics, 119334
Moscow, Russia}
 \author{Yu. Holovatch}\email[]{hol@icmp.lviv.ua}
 \affiliation{Institute for Condensed Matter Physics, National
Academy of Sciences of Ukraine,  UA--79011 Lviv, Ukraine}
\date{November 14, 2014}
\begin{abstract}
We discuss the non-self-averaging phenomena in the critical point of
weakly disordered Ising ferromagnet. In terms of the renormalized
replica Ginzburg-Landau Hamiltonian in dimensions $D <4$, we derive
an explicit expression for the probability distribution function
(PDF) of the critical free-energy fluctuations. In particular, using
known fixed-point values for the renormalized coupling parameters,
we obtain the universal curve for such PDF in the dimension $D=3$.
It is demonstrated that this function is strongly asymmetric: its
left tail is much slower than the right one.
\end{abstract}

\maketitle

\section{Introduction}

It is well known that the presence of weak quenched disorder in a
ferromagnetic system can essentially modify its critical properties
in the vicinity of the phase transition point such that new
universal critical exponents may set in
\cite{Harris74b,Khmelnitskii75,Lubensky75,Grinstein76,Dotsenko-book}.
On the other hand, in recent years it has been argued that due to
the presence of disorder the statistical properties of some
thermodynamical quantities at the critical point can become
non-self-averaging
\cite{NonSelfAverage1,NonSelfAverage2,NonSelfAverage3,RandFerro}.
The aim of the present study is to demonstrate that due to the
presence of weak disorder the statistics of the free energy
fluctuations in the critical point of the Ising ferromagnet is
described by a nontrivial universal distribution function.

Away from the critical point at scales much bigger than the
correlation length $R_{c}$ the situation is sufficiently simple:
here the system could be considered as a set of essentially
independent regions with the size $R_{c}$, and for that reason one
could naively expect that the free energy distribution function must
be Gaussian. In fact, besides the central Gaussian part (the
``body") this distribution has asymmetric and essentially
non-Gaussian tails \cite{RandFerro}. Approaching the critical point
one finds that the range of validity of the Gaussian body  shrinks
while the tails are getting of the same order as the body. Finally,
when the correlation length becomes of the order of the system size
(in the critical point) the free energy distribution function turns
into a universal curve.

Present investigation of the critical free energy fluctuations  is
performed in terms of the renormalized replica  Ginzburg-Landau
Hamiltonian in dimensions $D <4$, which allows us to derive the
explicit expression for their probability distribution function
(PDF). In particular, using known fixed-point values for the
renormalized coupling parameters, we obtain the universal curve for
such PDF in the dimension $D=3$ [Eq.(\ref{18}), Figure
\ref{figure1}].

\section{Renormalization group replica approach}

We consider the continuous version of the Ising ferromagnet in terms
of the random temperature $D$-dimensional Ginzburg-Landau (GL)
Hamiltonian:
\begin{eqnarray} \nonumber
H\bigl[\phi,\xi\bigr] &=& \int d^{D}{\bf x} \Bigl[\frac{1}{2}
\bigl(\nabla \phi({\bf x})\bigr)^{2} + \frac{1}{2} (\tau - \xi({\bf
x})) \phi^{2}({\bf x}) \\  \label{1} &+& \frac{1}{4} g \,
\phi^{4}({\bf x}) \Bigr]
\end{eqnarray}
where $\phi({\bf x})$ are scalar fields, $\tau = (T-T_{c})/T$ is the
dimensionless temperature parameter and $g$ is the usual GL coupling
parameter. The independent random quenched parameters $\xi({\bf x})$
are described by the Gaussian distribution  with $\overline{\xi({\bf
x})} = 0$ and $\overline{\xi^{2}({\bf x})} = 2u$ where the parameter
$u$ describes the strength of the disorder.

For a given realization of the disorder the partition function of
the considered system is
\begin{equation}\
 \label{2}
Z[\xi] \; = \; \int {\cal D} \phi \;
\exp\bigl(-H\bigl[\phi,\xi\bigr]\bigr) \; = \;
\exp\bigl(-F\bigl[\xi\bigr]\bigr)\, ,
\end{equation}
where $\int {\cal D}\phi$ denotes the integration over all
configurations of the fields $\phi({\bf x})$ and $F\bigl[\xi\bigr]$
is the (disorder realization dependent) free energy of the system.

The distribution function of the random free energy
$F\bigl[\xi\bigr]$ can be analyzed by studying the moments of the
partition function. Taking the integer $n$th power of the expression
in Eq.(\ref{2}) and performing the Gaussian averaging over the
disorder parameters $\xi({\bf x})$ we get the replica partition
function
\begin{equation}
 \label{3}
\overline{Z^{n}[\xi]} \; \equiv Z(n) \; = \int {\cal D} \phi_{1} ...
\int {\cal D} \phi_{n}
\exp\bigl(-H_{n}\bigl[\boldsymbol{\phi}\bigr]\bigr)\, ,
\end{equation}
where the replica Hamiltonian
\begin{eqnarray}\nonumber
H_{n}\bigl[\boldsymbol{\phi}\bigr] & = & \int d^{D}{\bf x} \Bigl[
\frac{1}{2} \sum_{a=1}^{n}\bigl(\nabla \phi_{a}({\bf x})\bigr)^{2} +
\frac{1}{2} \tau \sum_{a=1}^{n} \phi_{a}^{2}({\bf x}) \\ \label{4}
&+& \frac{1}{4} g \sum_{a=1}^{n}   \phi_{a}^{4}({\bf x}) -
\frac{1}{4} u \sum_{a,b=1}^{n}  \phi_{a}^{2}({\bf x})
\phi_{b}^{2}({\bf x})\Bigr]
\end{eqnarray}
depends on $n$ interacting fields $\boldsymbol{\phi} \equiv
\{\phi_{1}, ...\phi_{n}\}$.

Applying the renormalization group (RG) method to analyze the
Hamiltonian Eq. (\ref{4}), in dimensions $D = 4-\epsilon$ one does
not encounter in the one loop approximation the fixed point (FP)
with both non-zero coordinates $u^*\neq0$, $g^*\neq0$: this is
because the system of equations for the fixed points is degenerate
on the one--loop level \cite{Harris74b,Lubensky75,Grinstein76}. This
fixed point appears in the next, two--loop approximation. However,
the degeneracy of the one--loop equations leads to the
$\sqrt{\varepsilon}$--expansion \cite{Khmelnitskii75,Grinstein76}.
Being qualitatively correct, this expansion appears to be of no use
if the accurate quantitative results at $D=3$ are needed
\cite{Shalaev97,Folk99,Folk00}.

Alternatively, RG equations for the Hamiltonian Eq. (\ref{4}) have
been analyzed directly at $D=3$ using the minimal subtraction
\cite{Schloms} and massive \cite{Parisi} RG schemes. To evaluate the
divergent perturbation series in the renormalized couplings,
appropriate resummation technique has been used. Results of the
five-loop calculations based on the minimal subtraction scheme at
$D=3$ are given in Ref. \cite{Folk00}. In the massive RG scheme, the
most accurate results are obtained within accuracy of six loops in
Ref. \cite{Pelissetto00}. In particular, using two different
resummation schemes: based on the (i) conformal mapping and (ii)
Pad\'e approximants, the following estimates for the FP values were
obtained, respectively:
\begin{eqnarray}
\label{5}
\mbox{ (i): }&&  u_{*} \simeq 2.14, \hspace{2em} g_{*} \simeq 6.28, \\
\label{6} \mbox{ (ii): }&& u_{*} \simeq 1.98, \hspace{2em} g_{*}
\simeq 6.12,
\end{eqnarray}
cf. Eqs. (3.12) and (3.15) of Ref. \cite{Pelissetto00} (here,
instead of the notations for the coupling constants $\bar{u}^*$ and
$\bar{v}^*$
 of Ref. \cite{Pelissetto00}  the FP values (\ref{5}), (\ref{6})
 are given for the appropriately rescaled
renormalized couplings: $\pi*\bar{u}^{*}  \to u_{*}$ and
$\frac{8\pi}{9}\bar{v}^{*} \to g_{*}$).

Note, that these results for the 3D random Ising model stable FP
coordinates are far less accurate than those for the $O(m)$
symmetrical FP of the $m$-vector model. Further discussion and
comparison of contributions of different orders of perturbations
theory and interplay of different resummation schemes may be found
in Refs. \cite{Pelissetto00,Folk03}.

For further calculations of the critical free energy distribution
function we will take just the average of the two FP values
(\ref{5}), (\ref{6}) \cite{note1}:
\begin{equation}
u_{*} \simeq 2.06\, , \hspace{2em}  g_{*} \simeq 6.20\, . \label{7}
\end{equation}

\section{Critical free energy distribution function}

The idea of the further (somewhat heuristic) calculations of the
critical free energy distribution function is in the following.
According to the general approach of the RG theory of critical
phenomena in the vicinity of the phase transition point the total
free energy $F$ of the system can be decomposed into two essentially
different contributions:
\begin{equation}\label{7a}
 F \; = \; V f_{0} \; + \; V |\tau|^{2 - \alpha},
\end{equation}
where $V = L^{D}$ is the volume of the system ($L$ is its linear
size), $f_{0}$ is the {\em regular} (background) free energy density
(which remains finite and non-singular at $T=T_{c}$) and $\alpha$ is
the specific heat critical exponent. The second term ${\cal F} \; =
\; L^{D} |\tau|^{2 - \alpha}$ represents the fluctuating part of the
free energy which is singular at the critical point $\tau = 0$ and
it is this part that is calculated in terms of the RG theory. Taking
into account the standard relation among the critical exponents,
$D\nu = 2 - \alpha$ (where $\nu$ is the critical exponent of the
correlation length) one notes that {\it at} the critical point, when
the correlation length becomes of the order of the system size,
$R_{c} \sim |\tau|^{-\nu} \sim L$ the fluctuating part of the free
energy ${\cal F} \sim L^{(D\nu +\alpha -2)/\nu} \sim O(\ln L)$ is
getting {\em non extensive} with the volume of the system. It is the
distribution function of the  random quantity ${\cal F}$ in the
critical point that we are going to derive in this paper.

According to the general ideas of the RG  theory of critical
phenomena, in the vicinity of the critical point at small but
non-zero value of the temperature parameter $\tau$ the considered
system can be regarded as a set of $N \simeq V/R_{c}^{D}$
essentially independent ``cells" with the size of the order of the
correlation length $R_{c}$. The internal degrees of freedom of the
cells are integrated out, providing the renormalized FP values for
the coupling parameters $g \to g_{*}$ and $u\to u_{*}$
[Eq.(\ref{7})], as well as for the temperature parameter $\tau \to
\tau_{R} = \tau R_{c}^{1/\nu}$. Thus, the effective (renormalized)
Hamiltonian of the cell of the size $R_{c}$ is
\begin{equation}
\label{8} H_{R} \; = \; R_{c}^{D} \Biggl[ \frac{1}{2} \tau_{R}
\sum_{a=1}^{n} \phi_{a}^{2} + \frac{1}{4} g_{*} \sum_{a=1}^{n}
\phi_{a}^{4} - \frac{1}{4} u_{*} \sum_{a,b=1}^{n}  \phi_{a}^{2}
\phi_{b}^{2}\Biggr]\, .
\end{equation}
Correspondingly, in the critical point at $\tau=0$ and $R_{c} = L$
the partition function of the considered system can be estimated as:
\begin{eqnarray}\nonumber
Z(n) & \simeq & Z_{0}^{n} \, \prod_{a=1}^{n}
\Biggl[\int_{-\infty}^{+\infty} d\phi_{a} \Biggr] \exp\Biggl\{ L^{D}
\Biggl( -\frac{1}{4} g_{*} \sum_{a=1}^{n}   \phi_{a}^{4} \\
\label{9} &+& \frac{1}{4} u_{*} \sum_{a,b=1}^{n}  \phi_{a}^{2}
\phi_{b}^{2}\Biggr)\Biggr\}\, ,
\end{eqnarray}
where $Z_{0} = \exp\{- L^{D} f_{0}\}$ is the ``regular"
(non-singular) part of the partition function. Redefining $\phi_{a}
\to L^{-D/4} \, \phi_{a}$ instead of Eq.(\ref{9}) we get:
\begin{equation}
\label{10} Z(n) \; \simeq \; \exp\bigl\{ -n F_{0}(L)\bigr\} \,
\tilde{Z}(n)\, ,
\end{equation}
where
\begin{equation}\label{10a}
F_{0}(L) = L^{D} f_{0} + \frac{1}{4} D \ln L
\end{equation}
is the self-averaging part of the free energy and the reduced
partition function
\begin{eqnarray}\nonumber
\tilde{Z}(n) & = & \prod_{a=1}^{n} \Biggl[\int_{-\infty}^{+\infty}
d\phi_{a} \Biggr] \exp\Biggl\{ -\frac{1}{4} g_{*} \sum_{a=1}^{n}
\phi_{a}^{4}\\ \label{11} &+& \frac{1}{4} u_{*} \sum_{a,b=1}^{n}
\phi_{a}^{2} \phi_{b}^{2}\Biggr\}
\end{eqnarray}
defines the distribution function $P(f)$ of the finite
($L$-independent) fluctuating part of the free energy $f$ in the
critical point:
\begin{equation}
\label{12} \tilde{Z}(n) \; = \; \int_{-\infty}^{+\infty} df \, P(f)
\, \exp\{-n f\}\, .
\end{equation}
By definition, the fluctuating part of the free energy $f$ is the
mere difference between the total free energy $F$ and its
self-averaging part $F_0$, cf. Eqs. (\ref{7a}), (\ref{10a}):
\begin{equation}
\label{12a} f \; = \; F- F_0\, .
\end{equation}
By simple transformations the replica partition function,
Eq.(\ref{11}), can be represented as follows:
\begin{equation}
\label{13} \tilde{Z}(n) \; = \; \int_{-\infty}^{+\infty}
\frac{d\eta}{\sqrt{4\pi u_{*}}} \exp\Bigl\{-\frac{1}{4u_{*}} \,
\eta^{2}\Bigr\} \; G^{n}(\eta)
\end{equation}
where
\begin{equation}
\label{14} G(\eta) \; = \; \int_{-\infty}^{+\infty} d\phi \,
\exp\Bigl\{\frac{1}{2} \, \eta \, \phi^{2} - \frac{1}{4} g_{*}
\phi^{4} \Bigr\}\, .
\end{equation}
Performing analytic continuation from the integer  $n$ to arbitrary
complex values, according to eqs.(\ref{12}) and (\ref{13}), the
distribution function $P(f)$ can be obtained by the inverse Laplace
transform:
\begin{eqnarray}\nonumber
P(f) & = & \int_{-i\infty}^{+i\infty} \frac{ds}{2\pi i}
\int_{-\infty}^{+\infty} \frac{d\eta}{\sqrt{8\pi u_{*}}}
\exp\Bigl\{-\frac{1}{4u_{*}} \, \eta^{2} \; + \; s \, f\Bigr\} \\
\label{15} &\times& G^{s}(\eta)
\end{eqnarray}
or
\begin{equation}
\label{16} P(f) \; = \; \int_{-\infty}^{+\infty}
\frac{d\eta}{\sqrt{4\pi u_{*}}} \exp\Bigl\{-\frac{1}{4u_{*}} \,
\eta^{2} \Bigr\} \; \delta\Bigl(\ln[G(\eta)] \; + \; f\Bigr)\, .
\end{equation}
Introducing a new integration variable $t=\ln G(\eta)$ one gets
\begin{equation}
\label{17} P(f) \; = \; \int_{-\infty}^{+\infty}
\frac{dt}{\sqrt{4\pi u_{*}}} \; \frac{\exp\Bigl\{ t -
\frac{1}{4u_{*}} \, \eta^{2}(t) \Bigr\}}{G'[\eta(t)]} \; \delta(t +
f)\, .
\end{equation}
Thus, the final result for the free energy distribution function is
\begin{equation}
\label{18} P(f) \; = \; \frac{1}{\sqrt{4\pi u_{*}}} \;
\frac{\exp\Bigl\{ -f - \frac{1}{4u_{*}} \, \eta^{2}(-f)
\Bigr\}}{G'[\eta(-f)]}
\end{equation}
where the function $G(\eta)$ is defined in Eq.(\ref{14}), its
derivative
\begin{equation}
\label{19} G'(\eta) \; = \; \frac{1}{2} \int_{-\infty}^{+\infty}
d\phi \; \phi^{2} \; \exp\Bigl\{\frac{1}{2} \, \eta \, \phi^{2} -
\frac{1}{4} g_{*} \phi^{4} \Bigr\}
\end{equation}
and the function $\eta(-f)$ is defined by the equation
\begin{equation}
\label{20} \int_{-\infty}^{+\infty} d\phi \; \exp\Bigl\{\frac{1}{2}
\, \eta \, \phi^{2} - \frac{1}{4} g_{*} \phi^{4} \Bigr\} \; = \;
\exp\{-f\}\, .
\end{equation}
Here, the values of the FP couplings $g_{*}$ and $u_{*}$ are given
in Eq.(\ref{7}).

The universal curve for the probability distribution function
$P(f)$, Eq.(\ref{18}), is represented in Figure 1. We see that, like
in all the other systems where the free energy PDFs have been
calculated \cite{RandFerro,UnivRand,replicas} this function is
essentially non-symmetric: the left tail is much slower than the
right one.

\begin{figure}[h]
\includegraphics[scale=0.4,angle=00]{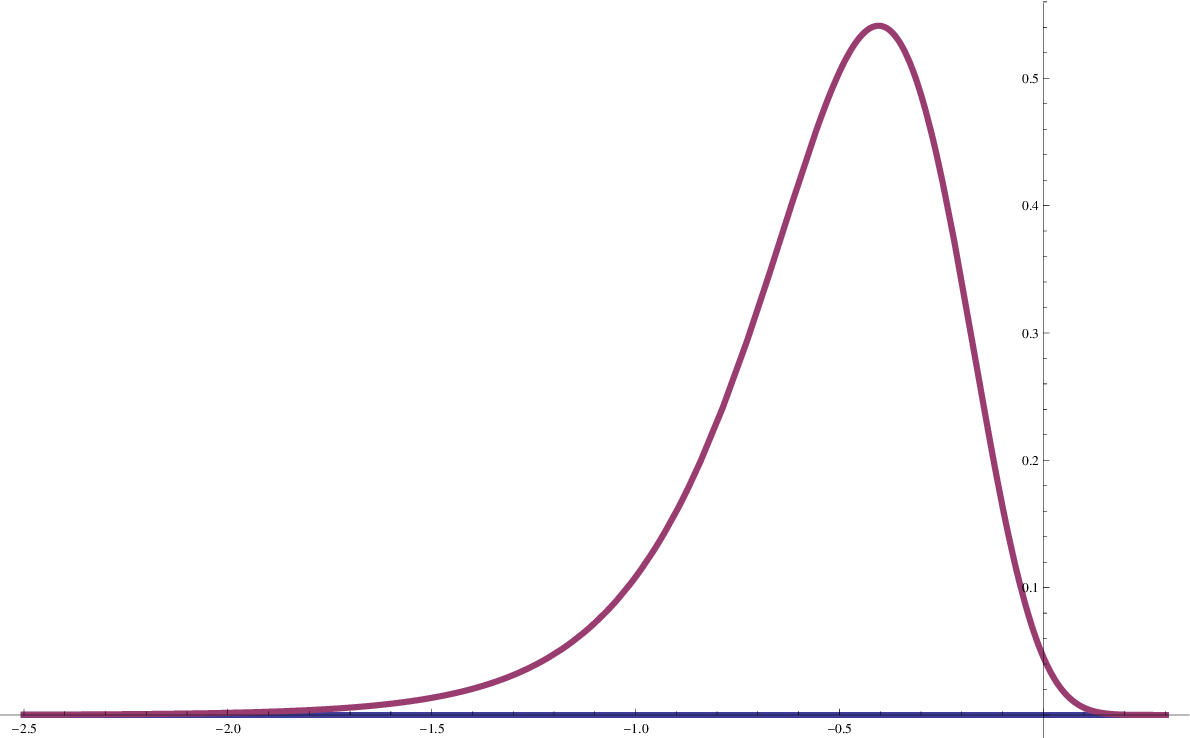}
\caption{(Color online) Critical free energy distribution function
of the disordered Ising ferromagnet in dimension $D=3$.}
\label{figure1}
\end{figure}

\subsection{Asymptotics}\label{IIIA}

Using Eqs.(\ref{18})-(\ref{20}) both the left and the right tails of
the probability distribution function $P(f)$ can be derived
explicitly.

In the limit $f \to -\infty$ the approximate solution of
Eq.(\ref{20}) is
\begin{equation}
\label{21} \eta(-f) \; \simeq \; \sqrt{4 g_{*} |f|}\, .
\end{equation}
Substituting this into Eqs.(\ref{19}) and (\ref{18}), and neglecting
pre-exponential factors one easily gets
\begin{equation}
\label{22} P(f\to -\infty) \; \sim \; \exp\Bigl\{ -
\frac{g_{*}}{u_{*}} \; |f| \Bigr\} \; \simeq \; \exp\Bigl\{ - 3.01
\; |f| \Bigr\}\, .
\end{equation}
In the opposite limit, $f \to +\infty$ the approximate solution of
Eq.(\ref{20}) is
\begin{equation}
\label{23} \eta(-f) \; \simeq \; -2\pi \; \exp\{ 2f \}\, .
\end{equation}
Substituting this solution into Eqs.(\ref{19}) and (\ref{18}), with
exponential accuracy one  gets
\begin{eqnarray}\nonumber
P(f\to +\infty) & \sim & \exp\Bigl\{ 2f - \frac{\pi^{2}}{u_{*}}
\exp\{ 4f \}\} \\ \label{24} & \simeq & \exp\Bigl\{ 2f - 4.79 \,
\exp\{ 4f \}\Bigr\}\, .
\end{eqnarray}
Note that this behavior is described by Gumbel distribution. Thus,
according to Eqs.(\ref{22}) and (\ref{24}), we see that the left
tail of the probability distribution function $P(f)$ is indeed much
slower than the right one.

\section{Conclusions}

In this paper we have derived explicit expression for the
probability distribution function of the free energy fluctuations of
weakly disordered three-dimensional Ising ferromagnet in the
critical point. First of all, it should be stressed that the mere
existence of such distribution function in the thermodynamic limit
means that the critical free energy fluctuations in the considered
system are {\em non-self-averaging}. This, of course is not
surprising as the values of these critical fluctuations are not
extensive with volume of the system. In this respect our analysis
differs from that of Refs.
\cite{NonSelfAverage1,NonSelfAverage2,NonSelfAverage3}, where
behaviour of extensive thermodynamic quantities at $T_c$ was
considered.

The other maybe more important result of the present research is
that obtained distribution function, Eqs.(\ref{18})-(\ref{20}) for
the fluctuating part of the free energy $f$, Eq.(\ref{12a}), Fig.
\ref{figure1}, is {\it universal}, which means that hopefully it
could be verified by e.g. numerical simulations. Of course it must
be not so easy to do, as the fluctuations under consideration must
be ``extracted" at the background of the leading extensive with the
volume self-averaging part of the free energy, but nevertheless we
hope that nothing is impossible for nowadays numerics...

\acknowledgments This work was supported by the International
Research Staff Exchange Scheme grant IRSES DCPA PhysBio-269139
within the Seventh Framework Program of the European Union. We thank
Gleb Oshanin for attracting our attention to the Gumbel distribution
in connection with the asymptotics described in Sec. \ref{IIIA}.


\begin{thebibliography}{99}
\bibitem{note1} In what follows we ignore possible dependence
of the FP values of the couplings on replica parameter $n$.
\end{thebibliography}
\end{document}